\documentclass[10pt,letterpaper]{article}

\usepackage{cogsci}

\cogscifinalcopy 

\usepackage{graphicx}
\usepackage{pslatex}
\usepackage{apacite}
\usepackage{subfig}
\usepackage{float} 
\usepackage{soul} 
\usepackage{color, xcolor} 

\title{Important Clues that Facilitate Visual Emergence: Three Psychological Experiments}
 
\author{{\large \bf Jingmeng Li (jmli17@fudan.edu.cn)} \\
  Laboratory of Algorithms for Cognitive Models, School of Computer Science, Fudan University, Shanghai, China
  \AND {\large \bf Hui Wei (weihui@fudan.edu.cn)} \\
  Laboratory of Algorithms for Cognitive Models, School of Computer Science, Fudan University, Shanghai, China}

\begin{document}

\maketitle

\begin{abstract}
  Visual emergence is the phenomenon in which the visual system obtains a holistic perception after grouping and reorganizing local signals. The picture \emph{Dalmatian dog} is known for its use in explaining visual emergence. This type of image, which consists of a set of discrete black speckles (speckles), is called an emerging image. Not everyone can find the dog in \emph{Dalmatian dog}, and among those who can, the time spent varies greatly. Although Gestalt theory summarizes perceptual organization into several principles, it remains ambiguous how these principles affect the perception of emerging images. This study, therefore, designed three psychological experiments to explore the factors that influence the perception of emerging images. In the first, we found that the density of speckles in the local area and the arrangements of some key speckles played a key role in the perception of an emerging case. We set parameters in the algorithm to characterize these two factors. We then automatically generated diversified emerging-test images (ETIs) through the algorithm and verified their effectiveness in two subsequent experiments.
\end{abstract}

\section{Introduction}
Perception is defined as the process of transforming signals from one's surroundings into the experience of objects, events, sounds, and tastes \cite{roth1986perception}. About 80\% of the information we receive each day comes from vision; this suggests that studying visual perception is an important way to explore human intelligence. We recognize words in books, objects on desks, or people in rooms so easily that we ignore the process from the initial visual signal to the emergence of perception. Studies have shown that about half of the cerebral cortex of primates participates in visual perception \cite{felleman1991distributed,dicarlo2012does}. Therefore, visual emergence has a high computational complexity.

\begin{figure}[h]
  \begin{center}
    \subfloat{\includegraphics[width=2in]{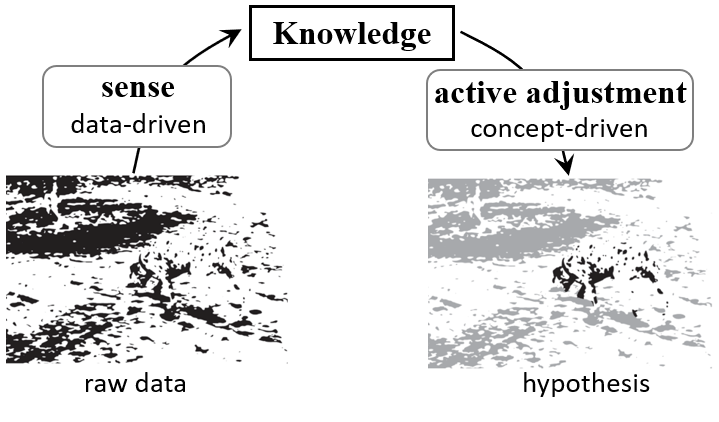}}
  \end{center}
  \caption{Explanation of the approximately perceptual process of the emerging image \emph{Dalmatian dog} in terms of information processing. The visual system perceives clues (edge and contour fragments) from speckles in the bottom-up process, and then it combines recognition clues and a priori knowledge to form cognitive hypotheses in the top-down process.}
  \label{figure1}
\end{figure}

Visual emergence is a phenomenon in which the visual system perceives meaningful wholes by integrating seemingly meaningless pieces \cite{holland2000emergence}. The picture \emph{Dalmatian dog} shown in Figure \ref{figure1} is often used to explain visual emergence. In the first half of the twentieth century, psychologists \cite{koffka2013principles,wertheimer1938gestalt} summarized the laws of perceptual processing and proposed Gestalt theory, which emphasizes the holistic nature of human perception and is the most widely accepted theory of perceptual organization. Discovering a dalmatian dog in the emerging image shown in Figure \ref{figure1} is an object recognition task. The whole process can be divided into two stages: bottom-up and top-down \cite{theeuwes2010top}. In the bottom-up process, the visual system groups and integrates physical signals according to gestalt principles and collects clues used for object recognition. In the top-down process, the visual system forms cognitive hypotheses by combining a priori knowledge and cognitive clues.

Although Gestalt theory offers several valuable principles with regard to grouping and reorganizing stimuli, it can neither demonstrate the adequacy and necessity of these grouping principles nor reveal how the visual system perceives a dalmatian dog from the emerging image. It is necessary, therefore, to explore what specific factors hinder or facilitate the occurrence of visual emergence. This not only has theoretical value for investigating visual perception in cognitive psychology but also helps promote the rational design of computer vision algorithms, thereby alleviating practices such as reliance on massive amounts of training data, expensive manual labeling, and huge computing power usage.

In object recognition, we use object features such as color, texture, and shape. Traditional object recognition theories emphasize that shape is more important in object recognition \cite{roberts1963machine}. Psychological-behavioral experiments have shown that surface information (color, texture) speeds up recognition but does not significantly improve recognition accuracy \cite{gegenfurtner2000sensory}. Biederman's recognition-by-components asserts that surface information only plays a role in low-level vision and provides cues for the organization and integration of visual signals while object recognition tasks rely on shape \cite{biederman1987recognition}. However, this view cannot explain discrimination between horses and zebras. If we only provide subjects with the shape of a zebra, they will likely mistake the zebra for a horse. The ``shape + surface'' computational framework for object recognition suggests that surface and shape information play a joint role in high-level visual processing, and that the role of surface information depends heavily on differences in structural properties between the objects in question \cite{tanaka2001role}.

According to ``shape + surface'' theory, the process of discovering the dalmatian dog in the emergent image shown in Figure \ref{figure1} can be described as follows. The visual system first obtains shape information, such as edges or contour segments, based on the physical features of visual signals in the bottom-up process. Then, it reorganizes the shape and surface information in the top-down process to discover more holistic combinations and form a cognitive hypothesis of a dalmatian dog based on a priori knowledge. Parsing the process backward gives us the following insight. Under the condition that a priori knowledge is available, the visual system reorganizes signals and collects clues in an iterative way. The results of clue collection affect the speed and accuracy of hypothesis formation, which in turn affects the speed and accuracy of finding the dalmatian dog in the emerging image. We believe, therefore, that the quality, quantity, distribution, and accessibility of recognition clues all have an effect on visual emergence. 

Emerging images differ from normal natural images in that they contain only black and white colors and no detailed texture information. Research on the visual cortex has found that neuronal cluster activity in primate areas V1 and V2 primarily reflects local luminance changes, and that neuronal activity in higher visual cortex areas represents global second-order features \cite{an2014orientation}. We infer, therefore, that the visual system appropriately uses some speckles to obtain recognition clues and also appropriately discards some speckles that might interfere. The criterion for the trade-off is whether a holistic hypothesis is promoted.

This study aimed to identify the factors influencing the occurrence of visual emergence using three psychological experiments. In the first, we recorded the behavior of subjects observing a typical emerging case. We analyzed the experimental data inductively to discover two factors that might affect visual emergence: the density of speckles (speckle-density), which affects the speed and accuracy of locating objects, and the arrangements of key speckles (speckle-arrangement), which contain discriminative texture and shape information that affect the accuracy of object recognition. We set parameters to describe these two factors in an algorithm and then automated the generation of emerging-test images (ETIs) using control variables. In the two subsequent experiments, we used these ETIs to verify the effectiveness of the two factors in influencing the occurrence of visual emergence.

\section{Experiment 1}
This experiment was designed to discover the factors that might influence the occurrence of visual emergence. Subjects were presented with a typical emerging case for them to observe. Their responses to it were recorded for analysis to generalize the factors that might have a significant effect on visual emergence.

\subsection{Participants}
A total of 120 students participated in this experiment (mean age = 22.8 years; 60 female). The participants come from the School of Computer Science, School of Psychology, School of Life Sciences, and School of Mathematics. None of the subjects had visual cognitive impairment.

\subsection{Procedure}
\begin{figure*}[h]
  \begin{center}
    \subfloat{\includegraphics[width=5.5in]{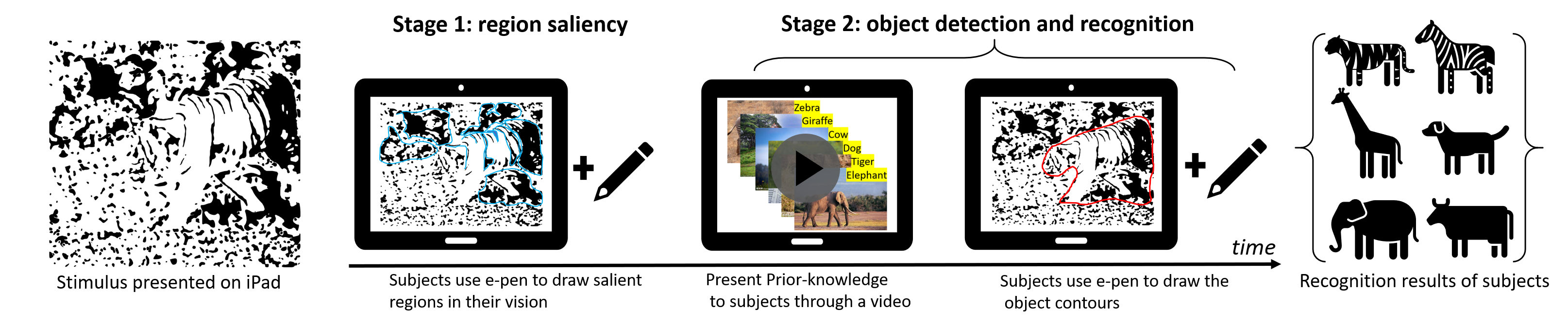}}
  \end{center}
  \caption{The flow of psychological experiment 1. It consisted of two stages. In stage 1, subjects were asked to circle regions according to the order in which they perceived them; in stage 2, subjects were asked to find and identify the animals in the emerging image based on the results drawn in stage 1.}
  \label{figure2}
\end{figure*}

Figure \ref{figure2} shows the flow of experiment 1. The whole process was divided into two stages. Subjects were required to perform the following operations using an iPad and an electronic pen (e-pen). In the first stage, the stimuli were presented on the iPad. The subjects observed them and then circled the corresponding regions in order of their perceived saliency with the e-pen. In the second stage, subjects were first presented with six pictures containing animals to ensure they had a priori knowledge of the object to be identified. Each image contained only one type of animal, and each was played for five seconds. Subjects were then asked to draw an outline of the object as completely as possible based on the range circled in the first stage and to identify which type of object it was.

\subsection{Results and Analysis}
\begin{figure}[h]
  \centering
  \subfloat[]{\includegraphics[width=1.13in]{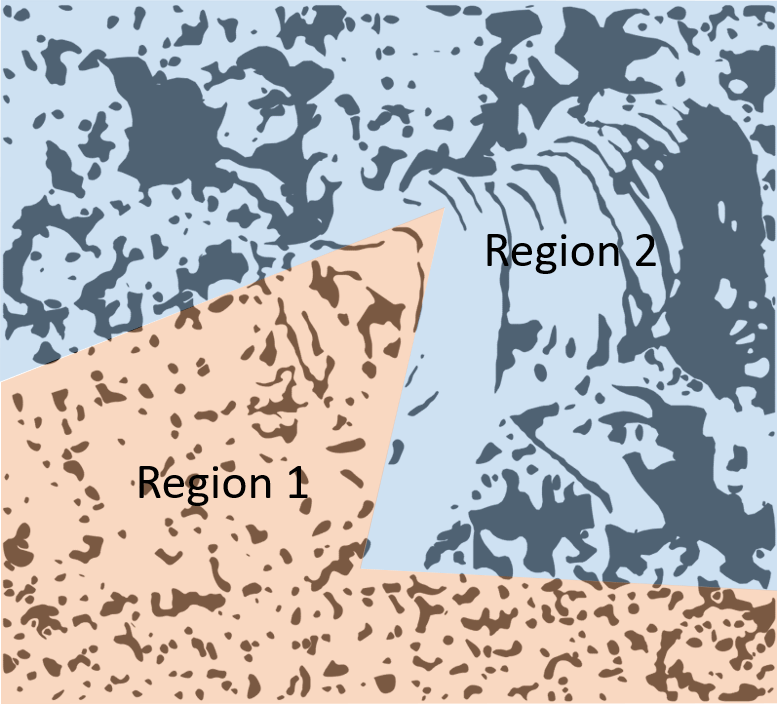}}
  \subfloat[]{\includegraphics[width=1.13in]{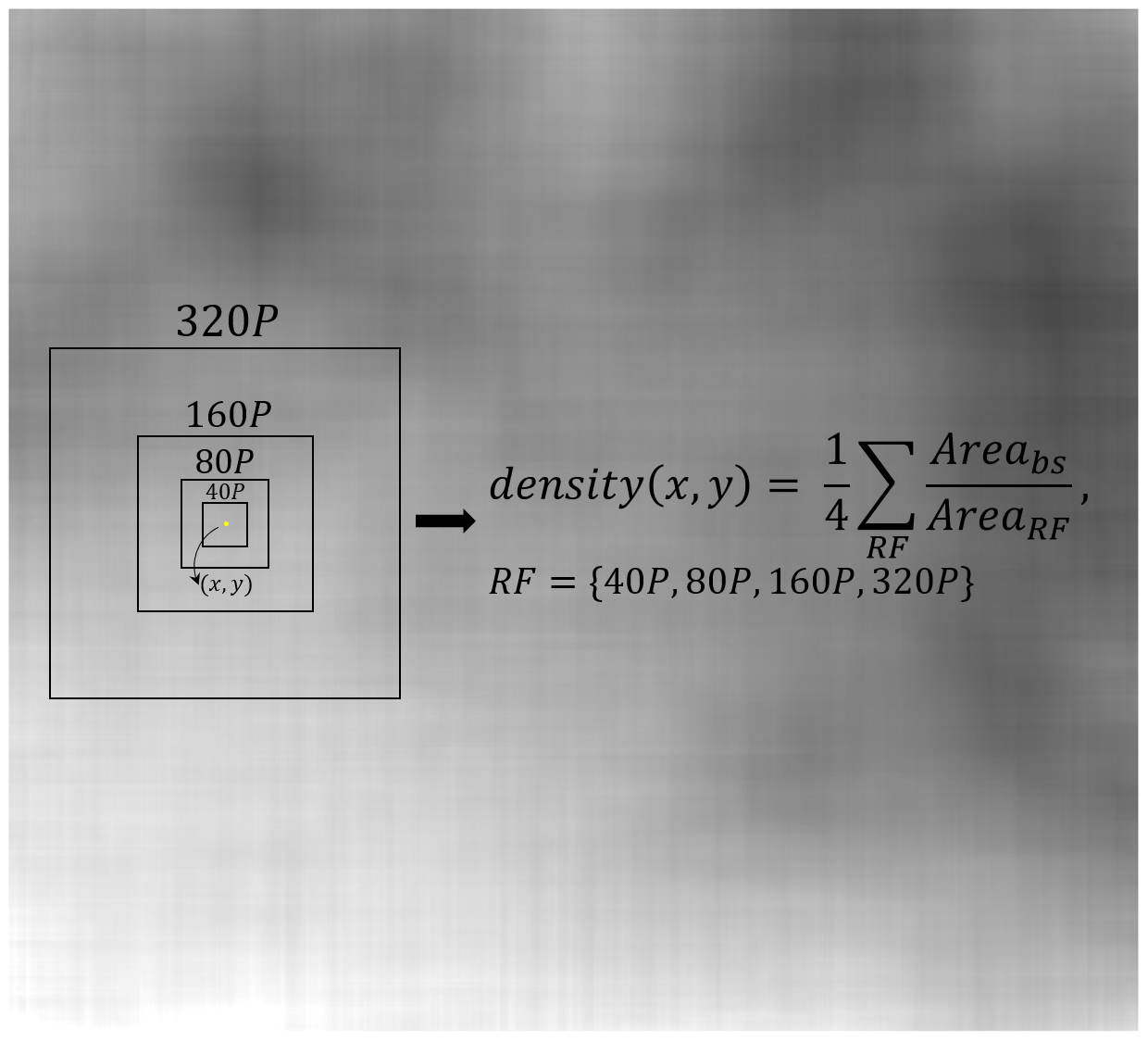}}
  \subfloat[]{\includegraphics[width=1.13in]{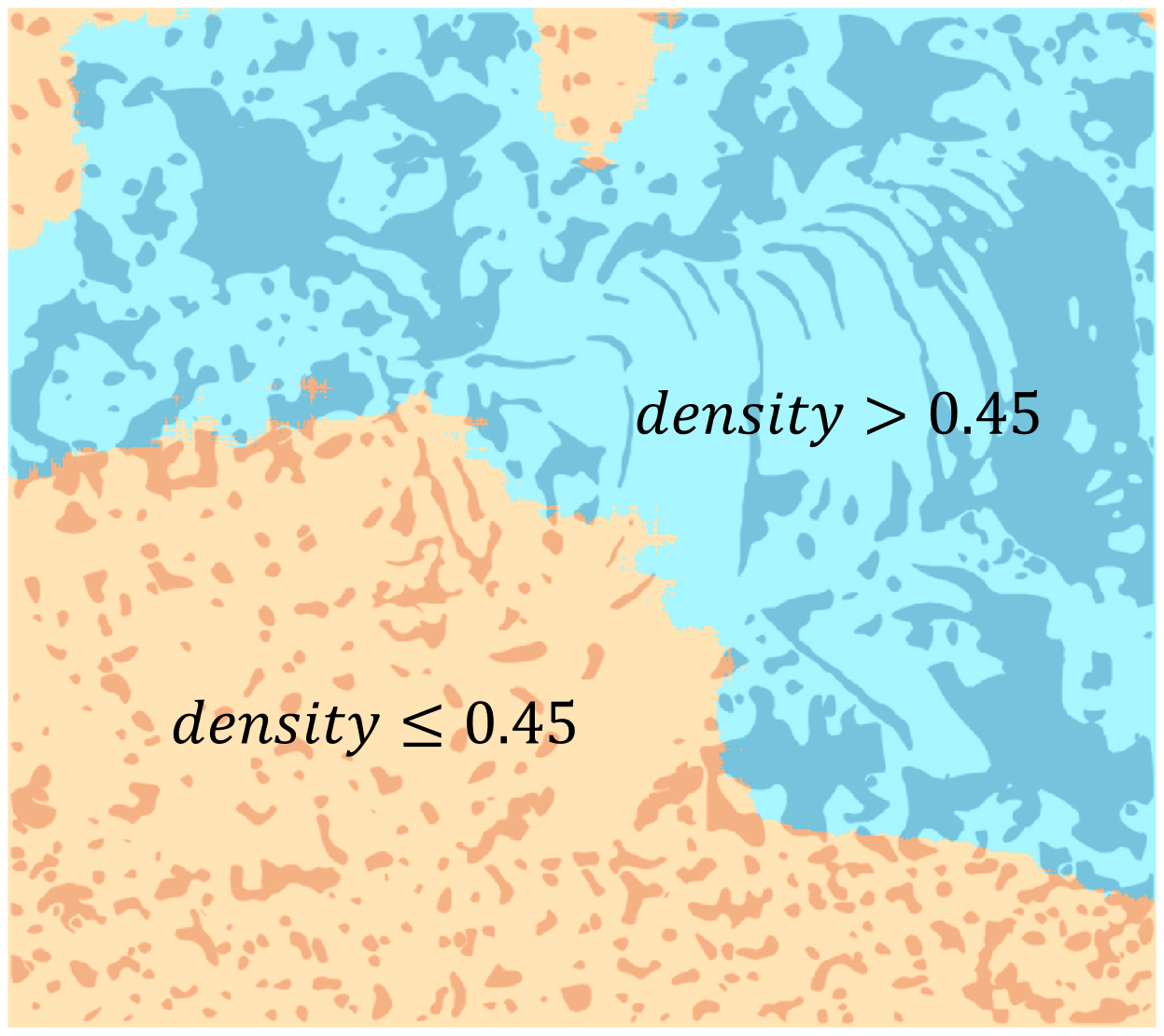}}
  \caption{Speckle-density might be an efficient clue; (a) shows the two regions divided based on the results drawn by the subjects; (b) explains the calculation of speckle-density and shows the normalized speckle-density map; (c) when we set the boundary of $density$ to 0.45, the emerging case is divided into two parts.}
  \label{fig3}
\end{figure}

The experimental data from stage 1 showed that the regions drawn by the subjects were not identical but highly overlapping, indicating that all subjects perceived the region as containing meaningful contours. Combining the results for all subjects, the emerging image can be roughly divided into Region 1 and Region 2, as shown in Figure \ref{fig3}(a). Region 1 covers 62.4\% of the whole image but contains 93.3\% of the contours drawn by the subjects. If the speckle-density in a local region is defined as the proportion of speckles to the total area of the region (i.e., $density\ =\ \frac{{\rm Area}_{bs}}{{\rm Area}_r}$), then the $density$ value of Region 1 is significantly larger compared with Region 2. Was it the speckle-density factor that made Region 1 more significant in the eyes of the subjects?

Neurophysiological studies have found that as the visual pathway deepens, the receptive fields (RF) of neurons gradually increase. In a setting with four different sizes of receptive fields, RF\ =\ {40 p, 80 p, 160 p, and 320 p}, where p denotes pixels, the speckle-density at each location is the mean value of $density$ at the four RFs. Figure \ref{fig3}(b) shows the normalized result of the emerging case after calculating the $density$ at all locations. If the boundary of $density$ is set to 0.45, the emerging case can be divided into two parts, as shown in Figure \ref{fig3}(c). Comparing (a) and (c) in Figure \ref{fig3}, we find that the region with $density>0.45$ in (c) overlaps well with Region 2 in (a). Therefore, speckle-density might significantly affect the occurrence of visual emergence.

According to ``shape+surface'' recognition theory, both shape and texture play a role in object recognition, and their importance depends on differences in the structural properties of the objects in question. Since color and its distribution is a texture feature, we only discuss the shape and texture information of the object. In this experiment, we provided the subjects with pictures of six animals as a priori knowledge---elephant, giraffe, cow, tiger, dog, and leopard. Among them, the elephant, giraffe, and tiger had large differences in shape and texture, and the dog, cow, leopard, and tiger had small differences in shape but large differences in texture. Therefore, texture is perhaps more effective than shape information for distinguishing tigers from other animals.

\begin{figure}[h]
  \centering
  \subfloat{\includegraphics[width=2in]{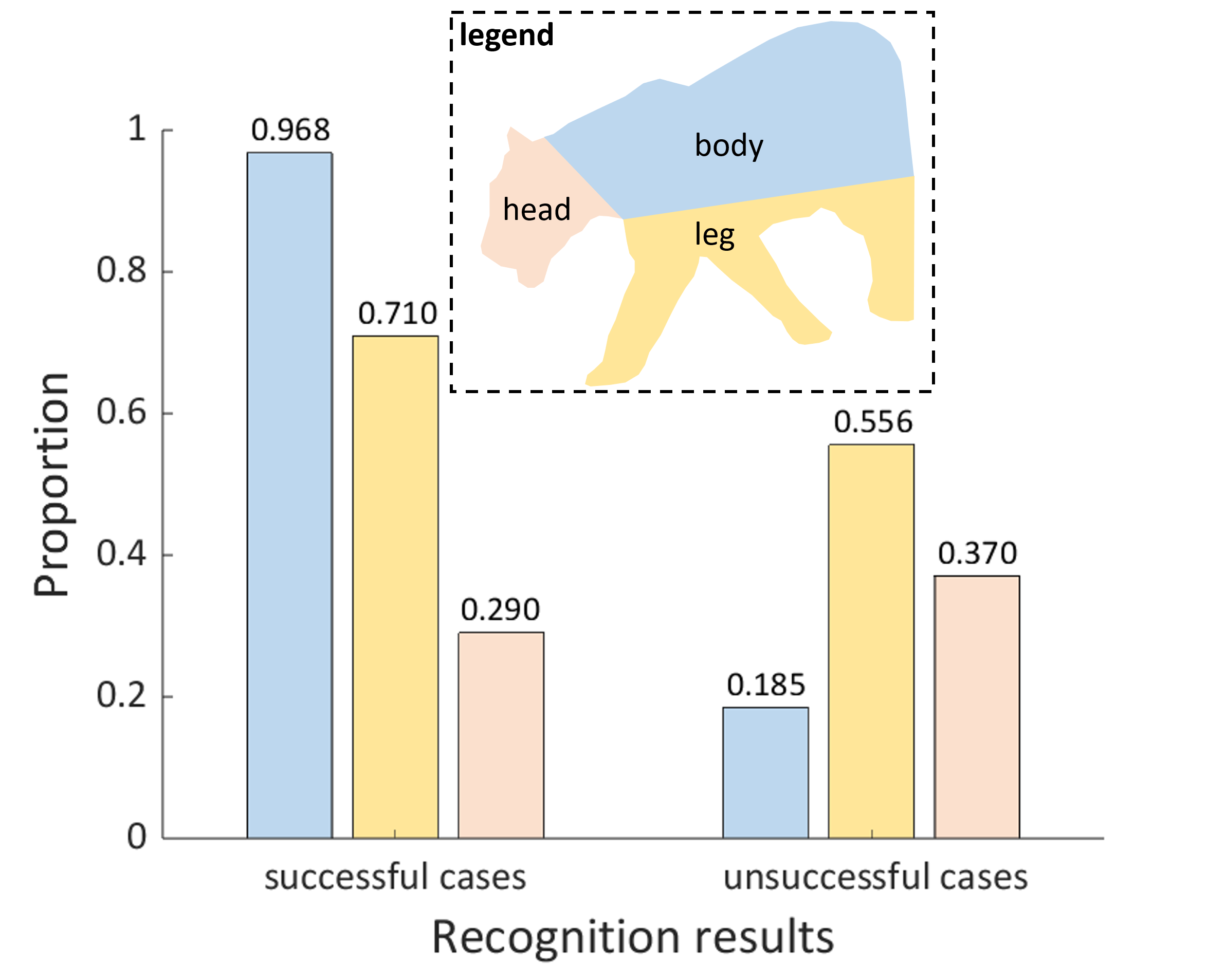}}
  \caption{Speckle-arrangement might be an important factor influencing the occurrence of visual emergence. Comparing the statistics of successful and unsuccessful cases, we can see that the texture feature in some part of the object plays a key role during perception. It is precisely the arrangement of speckles located on the body part that presents the texture feature.}
  \label{fig4}
\end{figure}

The experimental results of stage 2 showed that 93 of the 120 subjects successfully identified the tiger in the emerging case. Based on the results drawn by the subjects in stage 1, and combined with the object topology, the region containing the recognition clues can be divided into three parts---body, legs, and head---as shown in the legend of Figure \ref{fig4}. In the group of subjects who successfully identified the tiger, the percentages of subjects who correctly drew the body, legs, and head were 96.8\%, 71.0\%, and 29.0\%, respectively. In the group of subjects who did not successfully identify the tiger, the percentages of these three parts were 18.5\%, 55.6\%, and 37.0\%, respectively. Some subjects drew multiple parts of the recognition clues at the same time; thus, the statistics would show that the sum of the proportions of the three parts exceeded 1. We can conclude that the correct discovery of body and texture features was positively correlated with the successful identification of the tiger. The speckle-arrangement located on the body might have been responsible for the presentation of recognition clues. We suggest, therefore, that speckle-arrangement might affect the occurrence of visual emergence.

Based on the abovementioned experimental results and analysis, we identified two factors that might affect visual emergence: speckle-density and speckle-arrangement. When looking at the emerging image, we preferentially focus on the regions with greater speckle-density, and clues in these regions are more easily found. In addition, the quality of the recognition clues is critical to the final recognition. Occlusion often occurs between objects in complex environments, which can result in incomplete objects perceived by the visual system. The reason minor occlusions do not lead to false recognition is that high-quality clues, also called key clues, greatly facilitate recognition tasks. The texture presented by the arrangements of speckles on the body part is the key clue for identifying the tiger.

\section{Diversified ETI Generation}
It is unknown whether the two factors proposed in the first experiment are necessarily valid in the perception of emerging images. Therefore, their validity needed to be verified through corresponding psychological experiments. This required diversified ETIs in bulk. From an engineering standpoint, we can automate ETI generation with the help of a computer program with adjustable parameters---that is, controlling the values of the abovementioned two factors to be verified to automate the generation of stimuli under various settings.

\subsection{Natural Image Dataset}
We generated the corresponding ETIs based on natural images selected from the AM-2K dataset \cite{li2022bridging}. This dataset was created by Li et al. for natural image matting studies in computer vision, and it includes 2000 images in 20 animal categories. Most images in the AM-2K contain only one animal, and the positions and poses of the animals are rich and diverse, thus meeting the various needs of the second two experiments. In addition, the images in the dataset are high resolution, which makes data processing easier.

\subsection{Generation Process}
\begin{figure}[h]
  \centering
  \subfloat{\includegraphics[width=3.4in]{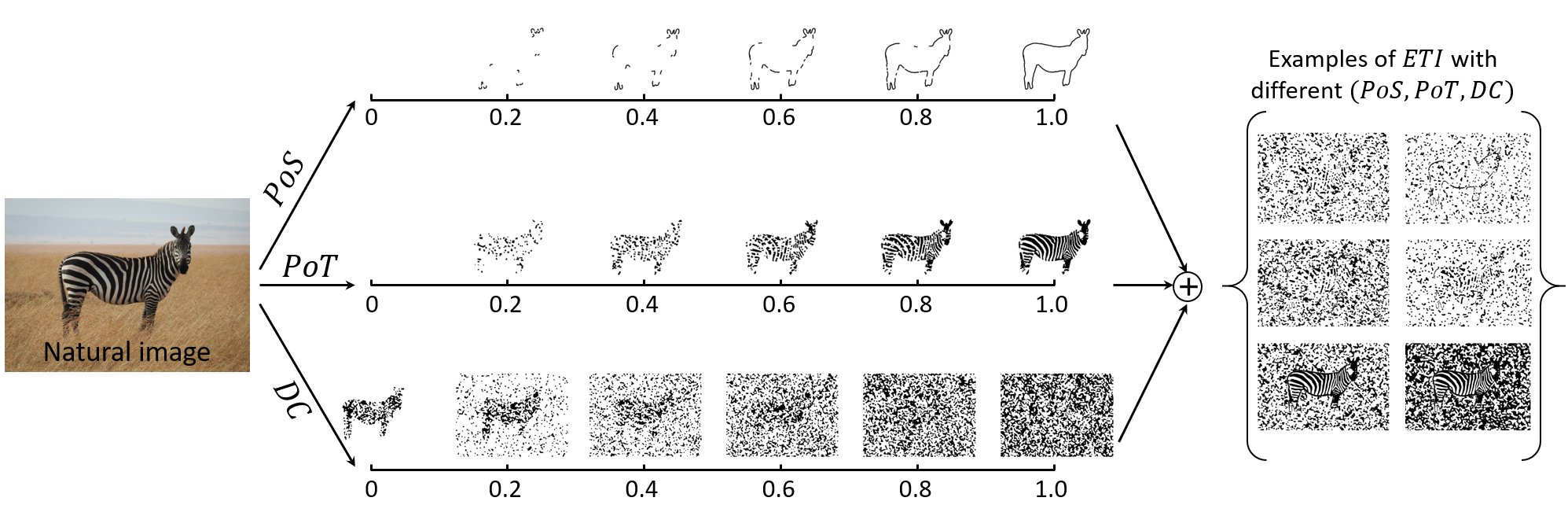}}
  \caption{Illustration of the method for generating diversified ETIs. Three parameters, $PoS$, $PoT$, and $DC$, are designed to control the diversity of testing samples. $PoS$ is used to adjust the proportion of shape, $PoT$ is used to adjust the proportion of texture information, and $DC$ is used to adjust the density of noise speckles around the object.}
  \label{fig5}
\end{figure}

To automate the efficient generation of ETIs similar to the emerging image \emph{Dalmatian dog}, the algorithm uses three parameters to characterize the two factors of speckle-density and speckle-arrangement. Figure \ref{fig5} uses a zebra image as an example to explain the generation process. Since both shape and texture can be used as key clues for object recognition, they are extracted as two independent dimensions. Object contours express the shape information; the parts of the contours with large curvature variations tend to include more specific shape information, and these contour segments are often more critical for discrimination. Thus, the normalized local curvature of the object contour was used to assess the importance of the contour segments. The first parameter, $PoS$, set in the program controls the proportion of contours rendered by speckles. For example, $PoS = 0.2$ means the first 20\% of the important parts of the contours are rendered by speckles in the ETI. The second parameter, $PoT$, controls the proportion of texture information. For example, $PoT = 0.2$ means 20\% of the texture will be randomly selected to be rendered by speckles in the ETI. When $PoS$ and $PoT$ are set, the speckle-density in the object region can be calculated. Then, the third parameter, the density contrast ($DC$), controls the density of noise speckles around the object. For example, $DC=0.2$ means the speckle-density of the surroundings is 20\% of the object region.

\section{Experiment 2a}
The purpose of this experiment was to test the validity of the first-factor speckle-density. Subjects were first presented with ETIs that reflected only changes in the parameter $DC$ to reduce the influence of other factors on the experimental results. We then verified its validity by analyzing the differences in the performance of subjects observing multiple groups of ETIs with different $DC$s.

\subsection{Participants}
The 120 subjects from Experiment 1 were invited to participate in this experiment because they already had some experience and could better cooperate with the experiment. In the experiment, the 120 subjects were divided equally into three groups: $G_1$ (mean age = 22.5 years), $G_2$ (mean age = 23.2 years), and $G_3$ (mean age = 22.7 years). The ratio of male to female in each group was kept the same as the overall ratio.

\subsection{Stimulus}
This experiment required subjects to perceive the region where the object was present from the ETIs. Therefore, the animals in the selected natural images had as much diversity as possible in terms of size, position, and pose to reduce the effect of visual habituation on the results. To avoid shape and texture interfering with the subjects' perceptions during the test, we set $PoS=0$ and $PoT=0$, and adjusted $DC=$ 0.2, 0.6, and 1, to generate the ETIs of 10 natural images. In the experiment, three ETIs corresponding to an image were presented to subjects in the three groups. For example, ${ETI}_{DC=0.2}^1$, ${ETI}_{DC=0.6}^1$, and ${ETI}_{DC=1}^1$, corresponding to the first natural image, were presented to subjects in the three subgroups, $G_1, G_2, and G_3$, respectively. In addition, a subject was presented with two successive ETI with different $DC$s to avoid visual habituation. For example, if the current subject was presented with ${ETI}_{DC=0.2}^1$, then the next subject was presented with ${ETI}_{DC=0.6}^2$ or ${ETI}_{DC=1}^2$.

\subsection{Procedure}
The stimuli presented in this experiment were generated by MATLAB Psychtoolbox for more accurate data collection. Subjects were seated 35 cm from a 24-inch $1920\times1080$ resolution monitor and given the following instructions: You are presented with a set of 10 ETIs, one at a time, with only one animal present in each ETI. Your task is to observe the currently presented ETI, and when you perceive the region of object presence from the ETI, draw a polygon with your mouse by clicking on the window to frame the region where you think the animal is present. When you are finished with the current ETI, click on the ``Next'' button and start to do the same for a new ETI.

\subsection{Results and Discussion}
During the experiment, the subject's reaction time (RT) in perceiving the object from the ETI was recorded. RT started from the presentation of the ETI until the subject drew the region where the object might have been present by clicking the window where the ETI was presented. This reflected the speed at which the subject perceived the object from the ETI. A smaller RT value indicated that the subject perceived the object from the ETI more easily. The time for an object to be perceived in an ETI was the average RT of 40 subjects.

\begin{figure}[h]
  \centering
  \subfloat[]{\includegraphics[width=1.6in]{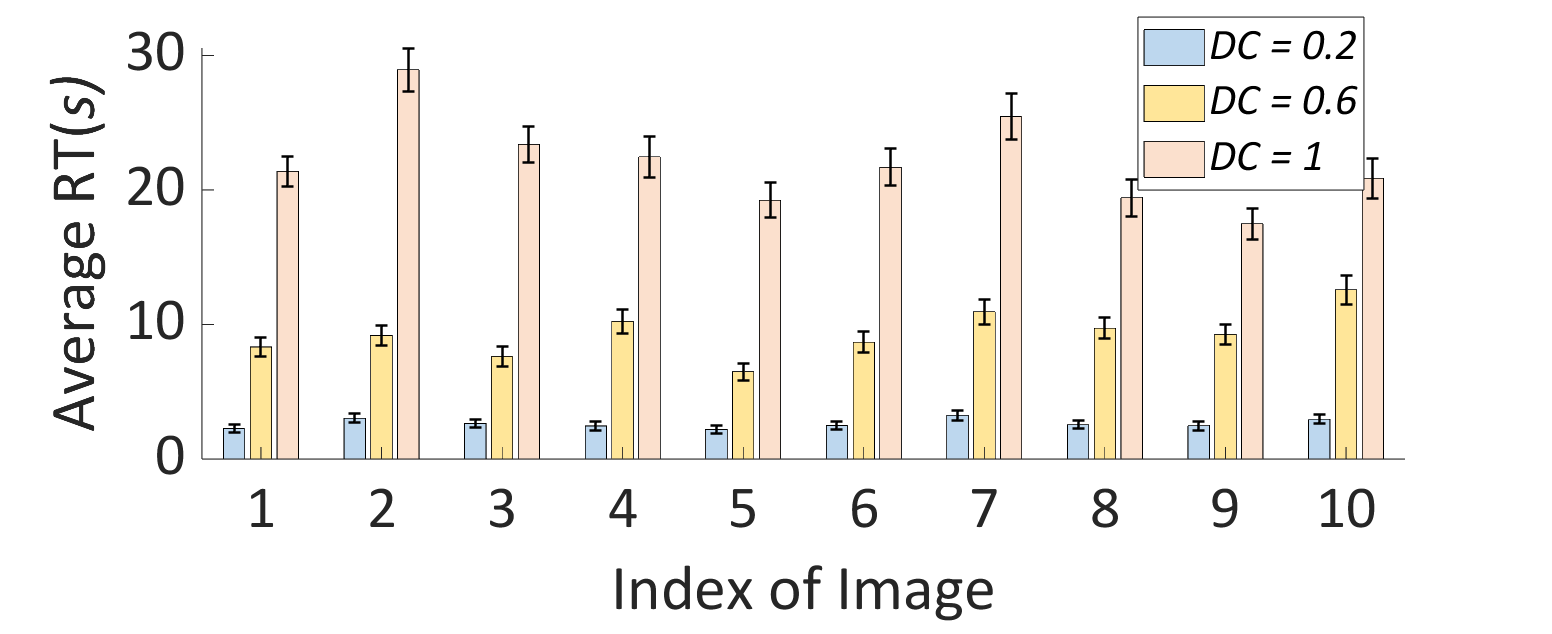}}
  \subfloat[]{\includegraphics[width=1.6in]{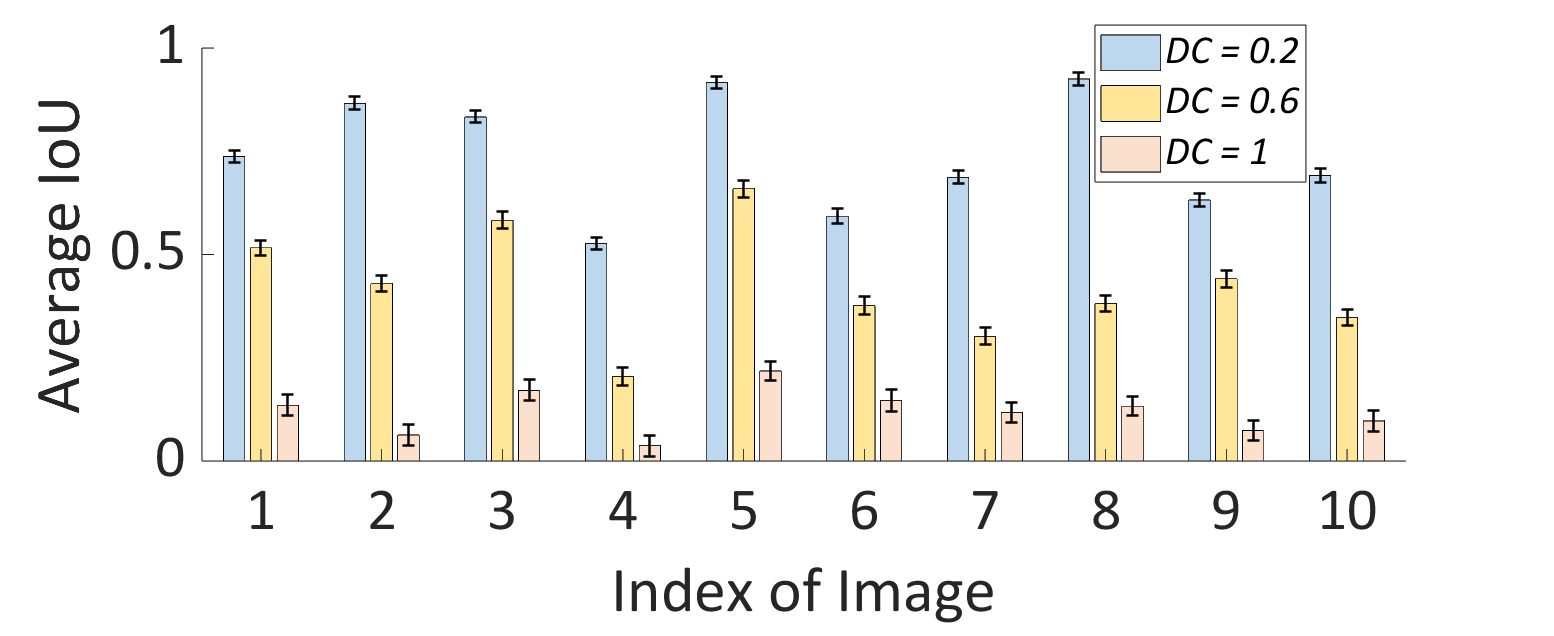}}
  \caption{Statistical results for subjects perceiving the object from ETIs with 95\% confidence intervals. (a) The strong positive correlation between the average RT and $DC$. (b) The strong negative correlation between the average IoU and $DC$. The results in (a) and (b) demonstrate that as $DC$ increased, subjects were more affected by interference speckles in the surroundings, resulting in a decrease in the speed and accuracy of perceiving the object.}
  \label{fig6}
\end{figure}

Figure \ref{fig6}(a) shows the average RTs of ETIs corresponding to 10 natural images under the parameter settings of $DC$ = 0.2, 0.6, and 1. The statistical results showed that although there were differences in the average RTs of ETIs with the same $DC$ for the 10 natural images, they all showed a trend of average $RT_{DC=0.2}$ $\textless$ average $RT_{DC=0.6}$ $\textless$ average $RT_{DC=1.0}$. Pearson correlation coefficient is often used to assess the strength of correlation between variables \cite{benesty2009pearson}. Pearson correlation coefficient, $r = 0.96$, between the average RT and $DC$ indicated a strong positive correlation between them. The experimental results demonstrated that a greater $DC$ means less interference speckles and more concentrated effective speckles, and visual emergence is more likely to occur. 

Intersection over union (IoU) was used to measure the accuracy of the object region framed out by the subjects. IoU is the ratio of the intersection and union between the region drawn by a subject and the actual region of the object; i.e., IoU = $\frac{{\rm Area}_s\cap{\rm Area}_o}{{\rm Area}_s\cup{\rm Area}_o}$. The region drawn by the subject is too large or too small to make IoU $\textless$ 1, and IoU = 1 only when the drawn region and the region where the object is located precisely coincide. Each ETI was finally obtained for 40 subjects. The accuracy of objects perceived in one ETI is the average IoU of 40 subjects. The closer the average IoU value is to 1, the more accurately the subject perceives the object from the ETI. Figure \ref{fig6}(b) shows the average IoUs of the ETIs with $DC$ = 0.2, 0.6, and 1 for 10 images. Pearson correlation coefficient between average IoU and $DC$ was -0.92, indicating a strong negative correlation between the two variables. The experimental results demonstrated that as $DC$ decreased, subjects' accuracy in perceiving objects from the ETI also decreased.

\section{Experiment 2b}
This experiment verified the second-factor speckle-arrangement. Subjects were first presented with ETIs reflecting only changes in speckle-arrangement. Then, the subjects observed them and identified the animals contained therein. The correlation between recognition accuracy and the parameters $PoS$ and $PoT$ can judge whether speckle-arrangement is effective for the occurrence of visual emergence.

\subsection{Participants}
We invited the same 120 subjects to participate in this experiment. The subjects were divided equally into two groups: $G_1$ (average age = 22.5 years) and $G_2$ (average age = 23.1 years). The proportion of male and female subjects in the two groups was the same as the overall proportion. We ensured that the 120 subjects had the required prior knowledge before the experiment.

\subsection{Stimulus}
When generating ETIs, the parameter $DC$ was set to 1 to avoid the influence of speckle-density. Then, $PoS$ and $PoT$ were sequentially adjusted to generate two sets of ETIs: ${ETI}_{PoT\ =\ 0.2,0.4,0.6,0.8,1}$ and ${ETI}_{PoS\ =\ 0.2,0.4,0.6,0.8,1}$. Since the subjects recognized objects based on the acquired shape and texture information, the shapes of the animals in the selected natural images were to remain intact and present a normal pose. We selected one image from the AM-2K for each of the six animals---tiger, zebra, leopard, camel, rhinoceros, and rabbit---to generate the corresponding two sets of ETIs. These two sets of ETIs were presented to the subjects in the order of progressively increasing parameters $PoS$ and $PoT$. In addition, two sets of ETIs of an image were presented to subjects in $G_1$ and $G_2$, and a group could not be presented with the same set of ETIs consecutively. For example, if the subjects in the two groups $G_1$ and $G_2$ were currently presented with ${ETI}_{PoT}^1$ and ${ETI}_{PoS}^1$, respectively, then the subjects in the next two groups would be presented with ${ETI}_{PoS}^2$ and ${ETI}_{PoT}^2$.

\subsection{Procedure}
The visual stimuli presented to the subjects in this experiment were generated by MATLAB Psychtoolbox. The subject sat 35 cm from a 24-inch 1920$\times$1080 resolution monitor and was instructed as follows: You are presented with a set of visual stimuli, one at a time, each containing only one animal and all containing the same animal in the same location. Your task is to look at the stimuli and determine what animals they contained. Click on the window when you are done to display the next stimulus, and then judge again.

\subsection{Results and Discussion}
In the experiment, the subjects' recognition results for each ETI were recorded. A correct recognition was recorded as 1; otherwise, 0. Each stimulus had 60 values of 0 or 1, and the mean value was the recognition accuracy, which reflected the difficulty of perceiving and recognizing the animals from ETIs.

\begin{figure}[h]
  \centering
  \subfloat[]{\includegraphics[width=1.4in]{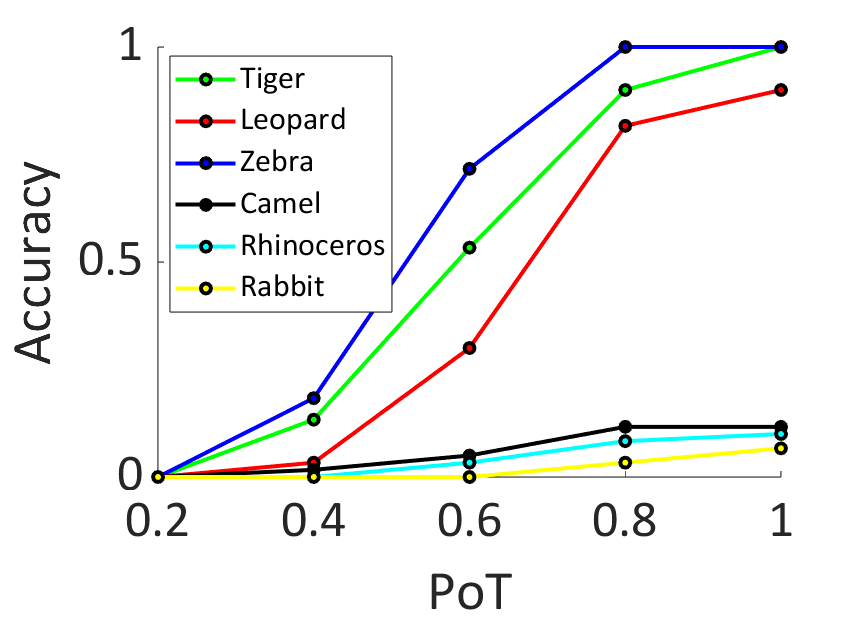}}
  \subfloat[]{\includegraphics[width=1.4in]{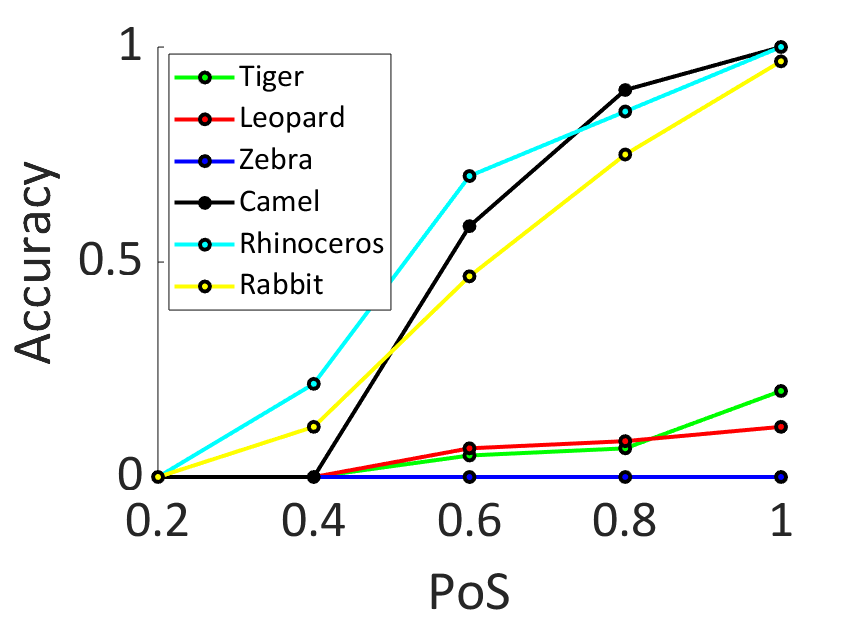}}
  \caption{Statistical results for the recognition accuracies for the six animals. (a) Changes in recognition accuracy for the six animals as $PoT$ was adjusted. (b) Changes in recognition accuracy for the six animals as $PoS$ was adjusted. Although $PoT$ and $PoS$ both affected recognition accuracies for the six animals, $PoT$ had a more pronounced effect for tiger, leopard, and zebra, and $PoS$ had a more pronounced effect for camel, rhinoceros, and rabbit.}
  \label{fig7}
\end{figure}

Figure \ref{fig7} shows the recognition accuracies of the ETIs corresponding to six animals with different $PoT$ and $PoS$ settings. The statistical results showed a significant positive correlation between speckle-arrangement and accuracy. However, the statistical results clearly reflected that varying the parameters $PoT$ and $PoS$ had different degrees of influence on accuracy for the six animals. The discriminative clue for tiger, leopard, and zebra is texture, while that for camel, rhinoceros, and rabbit is shape. Thus, adjusting $PoT$ resulted in more significant changes in the accuracies for tiger, leopard, and zebra, while adjusting $PoS$ resulted in more significant changes in the accuracies for camel, rhinoceros, and rabbit. The results of this experiment demonstrated that the speckle-arrangement is crucial for correct identification.

\section{General Discussion}
Speckle-density had a significant effect on the occurrence of visual emergence. Although Gestalt psychology explains the processing and organization of visual information in terms of several grouping principles, it is unclear how these principles function with emerging images. In fact, speckle-density and these principles are not contradictory. When the number of speckles in two equally sized regions is similar, the greater the speckle-density, the closer the speckles are to each other. If speckle-density continues to increase, then some speckles will overlap, which makes the object contour more continuous and complete. Among the Gestalt grouping principles, the process of speckle-density change exhibits proximity and continuity. Therefore, speckle-density is more accurate for explaining the occurrence of visual emergence.

We also found that some speckle-arrangements were important for accurate object recognition because they indirectly provided texture and contour information. This finding is instructive for current object recognition research in computer vision. Object recognition research based on deep networks has made considerable progress over the last decade. In terms of accuracy, some methods have even outperformed humans on some public datasets \cite{zhao2019object}. However, deep learning techniques rely heavily on the number of learned samples. By contrast, humans can easily recognize objects using only a small number of learning samples. Deep networks essentially rely on the denseness of the distribution of various samples to exhaust possibilities. Biological intelligence, meanwhile, cannot use resources so extravagantly, and biological brains learn more from interpretation \cite{ullman2007object,dicarlo2007untangling}.

Speckle-density and speckle-arrangement are factors that affect the occurrence of visual emergence, and the reason for this occurrence is the holistic precedence nature of human visual perception \cite{navon1977forest}. We perceive objects in terms of the global rather than the local, and even if part of the stimulus is altered, it does not affect the correct perception. By contrast, some studies have found that adding noise to images that could be correctly recognized or making changes that have no effect on human perception can lead to false recognition results. This can make deep networks unreliable for real-world applications (e.g., autonomous driving) and raise safety concerns \cite{li2021universal}. The abovementioned discussion implies that introducing certain biological mechanisms in the engineering domain could be very promising. The images for the emerging test discussed in this paper can help computer vision improve its ability to cope with unintended inputs.

\section*{Conclusion}
This study explored the factors that influence the perception of emerging images. The visual emergence process was divided into two stages---sense and recognition---and we separately examined the specific factors affecting the two parts. In the first experiment, we discovered two factors: speckle-density and speckle-arrangement. We automated the generation of ETIs in bulk with a computer program using a controlled-variable approach and then verified the effectiveness of two factors in the next two psychological experiments.



\bibliographystyle{apacite}

\setlength{\bibleftmargin}{.125in}
\setlength{\bibindent}{-\bibleftmargin}

\bibliography{CogSci_Template}

\end{document}